\documentclass[aps,pra,showpacs,twocolumn]{revtex4-1}
\usepackage{graphicx}
\usepackage[usenames]{color}
\usepackage{amssymb}

\newcommand{\fig}[1]{Fig.~\ref{#1}}
\newcommand{\be}[1]{\begin{equation}\label{#1}}
\newcommand{\ee}{\end{equation}}
\begin{document}

\title{Strongly driven molecules: probing the tunneling phase in the over-the-barrier regime and prevalence of different double ionization pathways}

\author{A. Emmanouilidou$^{1,2}$ and D. S. Tchitchekova$^{2}$}
\address{Department of Physics and Astronomy, University College London, Gower Street, London WC1E 6BT, United Kingdom $^{1}$\\
and Chemistry Department, University of Massachusetts at Amherst, Amherst, Massachusetts, 01003, U.S.A.$^{2}$
}

\begin{abstract}
Using a three-dimensional quasiclassical technique we explore
molecular double ionization by a linearly polarized, infrared
(800~nm) 27~fs laser pulse. For intensities ranging from the tunneling to the over-the-barrier regime,
we identify the double ionization pathways in a unified way as a function of total electron energy. For the tunneling regime, we discuss the differences 
 in the interplay of double ionization (DI) pathways between strongly driven He and strongly driven $N_{2}$. For intermediate intensities in the over-the-barrier regime, we find that both the correlated momenta and the double ionization probability distribution as a function of total energy probe the tunneling phase 
 of the re-colliding electron. This allows for a direct verification of the re-colliding electron tunneling at a large phase of the laser field in the over-the-barrier regime in contrast to a small tunneling phase in the tunneling regime. 
 
 \end{abstract}


 \maketitle

\section{Introduction}
Over the past two decades electron correlation has been established as
the underlying mechanism for many important phenomena arising from
the interaction of strong laser pulses with matter. One of these
phenomena is the dramatically enhanced multiple ionization yield of
atoms (e.g. \cite{WalkerPRL94}) and molecules (e.g.
\cite{AlnaserPRL2003}) in intense laser pulses.
When the field intensity is large the two electrons are stripped out sequentially \cite{lambropoulos1985} while for smaller intensities non-sequential double ionization (NSDI) dominates. 

According to the accepted mechanism for NSDI---the three-step or re-scattering model \cite{CorkumPRL93}---1) one electron escapes through the field-lowered Coulomb potential, 2) it moves in the strong infrared laser field and 3) it returns to the core (possibly multiple times) to transfer energy to the other electron in the parent atomic or molecular ion.
 Using coincidence imaging
techniques such as COLTRIMS many experiments have succeeded in
obtaining highly differential kinematical details of electron
correlation in the non-sequential intensity regime (e.g.
\cite{WeberNature2000,EreminaJPB2003,JesusJESRP2004,WeckenbrockPRL2004,StaudtePRL2007b,RudenkoPRL2007,LiuPRL2008}).
However, for higher intensities the available experimental data is
much less abundant. Correspondingly, the majority of theoretical
work has concentrated on the non-sequential intensity regime (e.g.
\cite{HoPRA2005,HaanPRL2006,BondarPRA2009}).

Addressing the double ionization of strongly driven systems
with first-principle quantum mechanical calculations in all three spatial
dimensions (3-d) is an immense task. Currently, 3-d ab-initio
quantum mechanical calculations are mostly available for the driven He
atom \cite{ParkerPRL2006}. With the exception of few  3-d ab-initio studies for diatomic molecules with fixed nuclei \cite{Bandrauk2},
 to cope with the highly complex task of
tackling double ionization of  molecules, many studies
use numerical quantum approaches of reduced dimensionality. Others use judiciously chosen quantum
models of reduced dimensionality (e.g. \cite{BaierPRA2008}), while
some use semi-analytical quantum approaches in the framework of the
so-called Strong-Field Approximation (SFA) (e.g.
\cite{FigueiraPRA2008}), not fully accounting for the Coulomb
singularity.

Here, we report on a classical study of electron correlation in the N$_2$
molecule driven by a 27 fs laser pulse at 800 nm for intensities well
within the non-sequential double ionization regime as well as for intensities corresponding to
the transition from pure tunneling to over-the-barrier ionization.
We use a three-dimensional quasiclassical technique that we developed for
conservative systems, i.e. the single photon multiple ionization of
atomic systems \cite{EmmanouilidouJPB2006,EmmanouilidouPRA2007,EmmanouilidouPRA2007b}.
We extended this technique to non-conservative systems to
first treat the correlated electron dynamics of the strongly driven He atom
\cite{EmmanouilidouPRA2008} and subsequently of the strongly driven $N_{2}$ molecule \cite{EmmanouilidouPRA2009}.  Our method is numerically
very efficient and treats the Coulomb singularity with no
approximation in contrast to techniques that use ``soft-core"
potentials.

We describe in a unified way the different double ionization (DI) pathways 
the two electrons follow to escape after re-collision as a function of total electron energy for the tunneling regime and for intensities corresponding to
the transition from pure tunneling to over-the-barrier ionization---we refer to the latter as intermediate intensities. (Throughout this work the total electron energy we refer to is the sum of the final 
kinetic energies of the two electrons). A similar description was very recently used to describe the DI pathways of strongly driven He \cite{EmmanouilidouNJP,EmmanouilidouPRA2010}.
This general treatment allows us to identify universal features of strongly-driven molecular double ionization and compare them with the atomic case.  In a previous study on strongly driven $N_{2}$ \cite{EmmanouilidouPRA2009}, we identified
the DI pathways only in the tunneling regime and independent of total electron energy for a very short 7 fs laser pulse. In our current study, we drive $N_{2}$ with a long laser pulse.  

We first discuss the general features of DI for different total energies and for intensities in the tunneling and the over-the-barrier regime. We then show that, for all intensities considered, the DI pathways have distinct
patterns of correlated momenta along the field polarization axis. 
Interestingly, for intermediate intensities in the over-the-barrier regime the correlated momenta of the Direct pathway---simultaneous ejection of both electrons---have an anticorrelation pattern. Moreover, the correlated momenta of the Delayed pathway---ejection of one electron with a delay of more that a quarter of a cycle with respect to re-collision---have a square pattern. Both the anticorrelation and square patterns arise from the ``soft" collisions, re-collision assisted field ionization \cite{EmmanouilidouPRA2009}, underlying the intermediate intensities in the over-the-barrier ionization. We show that the electron momentum corresponding to the boundary of the square pattern probes the tunneling phase of the re-colliding electron. 

Finally, we explore in detail the features of the DI probability distribution as a function of total electron energy. We find that in the tunneling regime the DI probability distribution has one peak, while for intermediate intensities in the over-the-barrier regime it has two peaks. We show that one peak---present for all intensities---is due to the field-assisted re-collision mechanism, and shifts to smaller energies with increasing intensity. The other peak is due to the ``soft" collisions underlying DI for intermediate intensities in the over-the-barrier regime. This latter peak probes in yet another way the tunneling phase. 




\section{Method}
 The method we use was previously described in \cite{EmmanouilidouPRA2009}. For completeness we also describe it in what follows.
Our 3-d quasiclassical model entails the following steps: We first
set-up the initial phase space distribution of the two ``active"
electrons in the N$_{2}$ diatomic molecule. Here, we consider only
parallel alignment between the molecular axis and the laser electric
field. At intensities in the tunneling regime we assume that one
electron tunnels through the field-lowered Coulomb potential. For
the tunneling rate one can use quantum mechanical or semiclassical
formulas for diatomic molecules (see, e.g.
\cite{TongPRA2002,LitvinyukPRL2003,KjeldsenJPB2004, LiPRA2007}). We
use the rate provided in ref. \cite{LiPRA2007}. The longitudinal
momentum is zero while the transverse one is provided by a Gaussian
distribution \cite{LiuPRL2007}. This description is valid as long as
the potential barrier is not completely suppressed by the
instantaneous laser field $E(t)=E_0(t) \cos (\omega t )$. We
consider the usual laser wavelength of 800~nm, corresponding to
$\omega=0.057\mathrm{a.u.}$ (a.u. - atomic units). In our
simulation the pulse envelope $E_{0}(t)$ is defined as
$E_0(t)=E_{0}$ for $0<t<7T$ and $E_{0}(t)=E_0 \cos^2(\omega
(t-7T)/12)$ for $7T<t<10T$ with T the period of the field. The
threshold for over-the-barrier ionization in neutral $N_{2}$, with an
ionization energy of $I_{p1}=0.5728$~a.u., is reached at a field
strength of $E = 0.075 \mathrm{a.u.}$ (corresponding to roughly
$2\times10^{14} \mathrm{Watts/cm}^2$).

Above $2\times10^{14} \mathrm{Watts/cm}^2$ the laser field allows an
unhindered electron escape and therefore the initial phase space is
modeled by a double electron microcanonical distribution
\cite{MengPRA89}. However, in
setting-up the initial phase space distribution we transition
smoothly from the tunneling to the over-the-barrier intensity
regime. Namely, we assign a random number to the phase $\phi$ of the
laser field when the first electron is ionized, see
\cite{BrabecPRA96,YePRL2008}. If the phase $\phi$ corresponds to an
instantaneous strength of the laser field $E(\phi)$ that leaves the
electron below the barrier then we use the initial conditions
dictated by the tunneling model. If the instantaneous field strength
pushes the barrier below the $I_{p1}$ of that electron then we use
the microcanonical distribution to set-up the initial phase space
distribution. This choice of initial conditions has proven
successful in past studies \cite{YePRL2008} in modeling the
experimental ratio of double versus single ionization for long laser
pulses  \cite{CornaggiaPRA2000}. With our approach we ensure a
smooth transition of the initial phase space distribution as we
change the intensity. Even at an intensity of
3$\times10^{14}\mathrm{Watts/cm^{2}}$ still about 70\% of the double
ionization probability corresponds to trajectories initialized using
the tunneling model, while 30\% of the probability corresponds to
trajectories initialized using the microcanonical distribution. 

After setting-up the initial phase space distribution we transform
to a new system of coordinates, the so called ``regularized"
coordinates \cite{KustaanheimoJRAM65}. This transformation is exact
and explicitly eliminates the Coulomb singularity. This step is more
challenging for molecular systems since one has to ``regularize"
with respect to more than one atomic centers versus one atomic
center for atoms.  We regularize using the global regularization
scheme described in ref.~\cite{HeggieCelMech74}. Finally, we use the
Classical Trajectory Monte Carlo (CTMC) method for the time
propagation \cite{AbrinesProcPhysSoc66c}. The propagation involves
the 3-d four-body Hamiltonian in the laser field with ``frozen"
nuclei:
\begin{eqnarray*}
    H = \sum_{i=1}^{2} [\frac{p_{i}^{2}}{2}-\frac{1}{|\vec{R}/2-\vec{r}_{i}|}-\frac{1}{|-\vec{R}/2-\vec{r}_{i}|}]\\
      +\frac{1}{|\vec{r}_{1}-\vec{r}_{2}|}+(\vec{r}_{1}+\vec{r}_{2}) \cdot \vec{E}(t),
\end{eqnarray*}
where $E(t)$ is the laser electric field polarized along the
z direction and further defined as detailed above, and $\vec{R}$ is
the internuclear distance.

\section{Results and discussion}
We consider three laser intensities at 10$^{14}$, 1.5 $\times 10^{14}$ and 3 $\times 10^{14}$ Watts/cm$^2$.
The range of laser intensities we consider is important because, at
800 nm, for 1.44$\times$ 10$^{14}$ Watts/cm$^2$ the maximum return
energy of the re-colliding electron (3.2 U$_{p}$ according to
the three-step model \cite{CorkumPRL93}) equals the first ionization energy of the ground state of $N_{2}^{+}$. The ponderomotive energy
$U_{p} = E_{0}^2/(4\omega^2$) is the cycle-averaged energy of the oscillatory motion.

\subsection{Mean values}    
\begin{figure}
\centering
\includegraphics [scale=0.4] {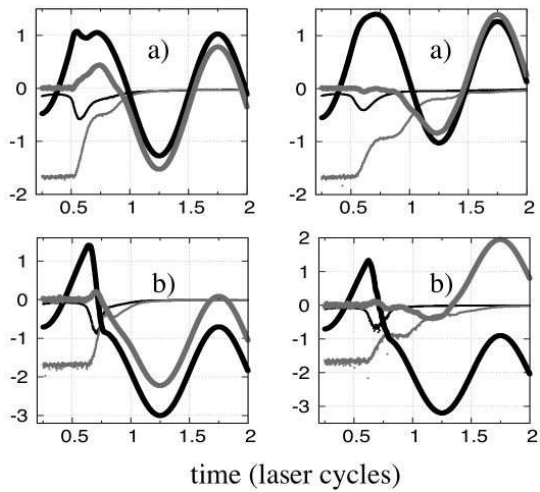}
\caption{We plot $\left <p_{1,z} \right>$ (thick black line), $\left<p_{2,z}\right>$ (thick grey line), $\left< -Z/|\vec{R}/2-\vec{r}_{1}|-Z/|-\vec{R}/2-\vec{r}_{1}| \right>$ (thin black line) and $\left<-Z/|\vec{R}/2-\vec{r}_{2}|-Z/|-\vec{R}/2-\vec{r}_{2}|\right>$(thin grey line). The time axis
is measured in laser cycles. Electron 1 is the re-colliding electron, $\vec{r}_{i}$ is the position vector of electron $i$, and $Z=1$.  The plotted quantities are shown for  1.5 $\times 10^{14}$ Watts/cm$^2$: top row for energies [0,1]$U_p$,  bottom row for energies [8,9]$U_p$. Left column refers to the SE pathway and right column to the RESIb pathway.}
\label{mean1513}
\end{figure}

\begin{figure} 
\centering 
\includegraphics [scale=0.35] {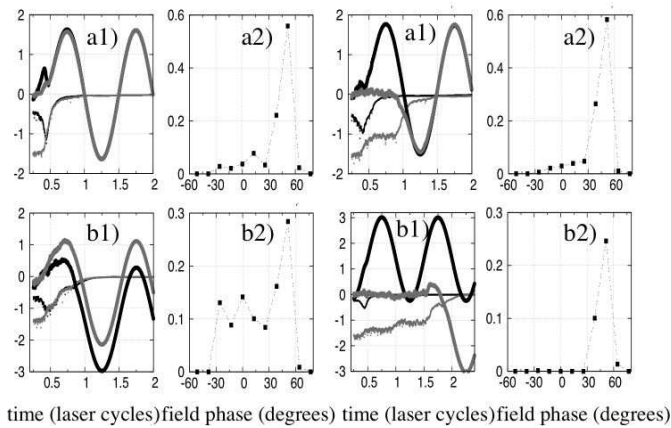} 
\caption{First and third columns from the left same as \fig{mean1513} but for 3 $\times 10^{14}$ Watts/cm$^2$: top row for energies [0,0.4]$U_p$, bottom row for energies [3.2, 3.6]$U_p$. We also show the probability distribution for the phase when the re-colliding electron tunnels out for the SE pathway, second column from the left, and for the RESIb pathway, fourth column from the left.  }
\label{mean3013}
\end{figure}
In this section, we briefly explore the characteristics and general properties of the double ionization (DI) pathways. Our goal is to point out that for different intensities and different total energy regimes the DI pathways have different characteristic features.

To identify the main DI energy transfer pathways we use the time delay between the re-collision time and the time of ionization of each electron \cite{EmmanouilidouPRA2009}. We define the re-collision time as the time of minimum approach of the two electrons. We identify this time through the maximum in the electron pair potential energy. The ionization time for each electron is defined as the time when the sum of the electron's kinetic energy (using the canonical momentum) and potential energy with the two atomic centers becomes positive and remains positive thereafter---for more details for the time of ionization see \cite{EmmanouilidouPRA2008} and references there in.  

Following the steps outlined above, we identify the following pathways: the Direct (SE) and Delayed (RESI) which are well established. (The delayed pathway is also referred to as re-collision-induced excitation with subsequent field ionization, RESI \cite{KopoldPRL2000,FeuersteinPRL2001}). In the direct ionization pathway (SE) both electrons are
ionized simultaneously very close (less than a quarter of a laser period) to the re-collision time. In the delayed ionization pathway, the re-colliding electron excites the remaining electron but does not ionize it. The electron is subsequently ionized at a peak (RESIa) or at a zero (RESIb) of the laser electric field. In addition to the SE and RESI pathways we also identify the double delayed pathway (DDE) with both electrons ionizing more than a quarter of a laser cycle after re-collision. This pathway was first identified in our recent study of strongly driven He \cite{EmmanouilidouPRA2010}.
 
 In \fig{mean1513} for 1.5$\times$ 10$^{14}$ Watts/cm$^2$ and in \fig{mean3013} for 3$\times$ 10$^{14}$ Watts/cm$^2$,
we plot the average of the momentum component of each electron along the
polarization axis (the z-axis), $<p_{1,z}>$ and $<p_{2,z}>$, for the SE and RESIb pathways. In addition, we plot  the average of the potential energy of each electron with the two nuclei.  
The time of re-collision is identified in \fig{mean1513} and \fig{mean3013} as the time at which there is a sudden rise/dip of the
nuclear contribution to the potential energy of the second/first electron. In the SE
pathway, electron 2 ionizes at a time close to the time of re-collision while in RESIb
electron 2 ionizes around T/2 later. In both \fig{mean1513} and \fig{mean3013} we consider only those
trajectories where the re-collision occurs at the first return of the re-colliding electron
to the atomic centers. For 800 nm, we find this to be the most important contribution to the
SE and RESI pathways for all intensities.
Nevertheless, multiple returns of the re-colliding electron are explicitly accounted for in the rest of our results. In addition, the general properties of SE and RESIb pathways for one return
of the re-colliding electron, described below, hold true for multiple returns the only difference being the re-collision time.

{\it Tunneling regime.} For intensities in the tunneling regime above 1.44$\times$ 10$^{14}$ Watts/cm$^2$, 
we find that the re-collision time shifts from T/2 for small total energy to (2/3) T (the time of
maximum energy re-collision in the three-step model) for large total energy---see \fig{mean1513} and compare the re-collision times of a) and b).  This is similar to our finding for strongly driven He for intensities where 3.2 $U_{p}$ is above the first excitation energy of He$^{+}$ \cite{EmmanouilidouNJP}.
 The behavior
of  $<p_{1,z}>$ against time also differs markedly as we go from one energy regime to the next. For small total energy, 
  at the instant of re-collision the momentum of the re-colliding electron first decreases due to transfer of energy to 
electron 2---dip in $<p_{1,z}>$ around T/2, see the SE pathway depicted on the top row of \fig{mean1513}. But at this early re-collision
time the laser field does not undergo a sign change and so continues to pull electron 1 in
the same direction as prior to re-collision. For high total energy,  
 the interaction of the re-colliding electron with the nucleus becomes very strong. At 2/3T (re-collision time at high total energies) electron 1 gets sharply pulled back both by the laser field and the nucleus and reverses in direction, see bottom row 
 of \fig{mean1513}.

{\it Over-the-barrier regime.} For intermediate intensities in the over-the-barrier regime, for the majority of trajectories, re-collision takes place early at T/2, see \fig{mean3013}. Moreover, as shown in \fig{mean3013}, we find that the re-colliding electron for the majority of DI trajectories tunnels when the phase of the laser field is around 50$^{\circ}$---much larger than 17$^{\circ}$, the phase that according to the three-step-model corresponds to maximum return energy. At 3$\times$ 10$^{14}$ Watts/cm$^2$, tunneling at a large phase of the field around 50$^{\circ}$ results in a maximum excursion of electron 1 in the laser field, before its return to the molecular ion, which is smaller than the maximum excursion at 1.5$\times$ 10$^{14}$ Watts/cm$^2$---small tunneling phase in the tunneling regime. (That larger tunneling phase results in a smaller maximum excursion can be easily seen using the simplest version of the three-step model.)  Indeed, focusing on the potential energy of electron 1 in \fig{mean3013} and \fig{mean1513}, we find that at T/4 the potential energy is smaller at 3$\times$ 10$^{14}$ Watts/cm$^2$ than at 1.5$\times$ 10$^{14}$ Watts/cm$^2$. Moreover, upon its return to the molecular ion, electron 1 transfers a smaller amount of energy to electron 2 at 3$\times$ 10$^{14}$ Watts/cm$^2$ than at 1.5$\times$ 10$^{14}$ Watts/cm$^2$. Indeed, the change in the potential energy of electron 1 during re-collision is smaller at 3$\times$ 10$^{14}$ Watts/cm$^2$ than at 1.5$\times$ 10$^{14}$ Watts/cm$^2$, compare \fig{mean3013} and \fig{mean1513}. The fact that the two electrons undergo ``soft" collisions at 3$\times$ 10$^{14}$ Watts/cm$^2$ is further supported by our finding that the re-colliding electron ionizes before the re-collision time for 30\% of the SE trajectories and for 70\% of the RESI ones.  Thus, in contrast to ``soft" collisions underlying DI for intermediate intensities in the over-the-barrier regime, it is strong collisions with both electrons ionizing after re-collision that underly DI in the tunneling regime.


\subsection{Correlated Momenta} 
In this section, we further explore the presence of ``soft" collisions at an intensity 3 $\times$ 10$^{14}$ Watts/cm$^{2}$ and identify their traces on the correlated momenta for each DI pathway. 

\begin{figure} [h]
\centering 
\includegraphics [scale=0.25] {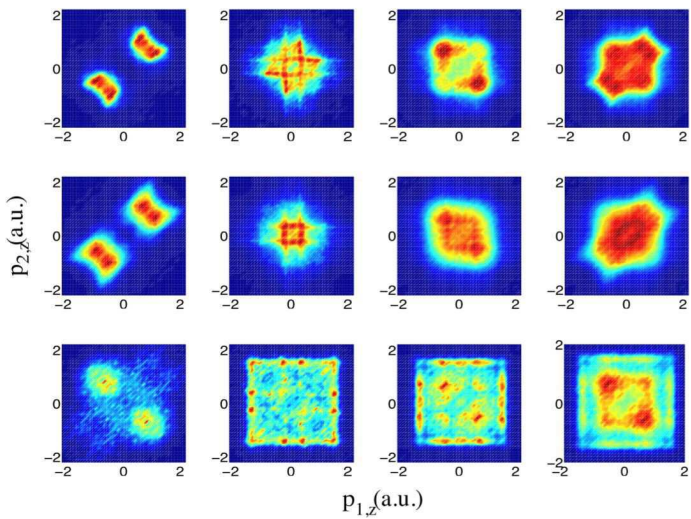} 
\caption{Correlated momenta at intensities $10^{14}$ Watts/cm$^2$ (top row), $1.5 \times 10^{14}$ Watts/cm$^2$ (middle row), and $3 \times 10^{14}$ Watts/cm$^2$ (bottom row). Columns from left to right indicate the various DI mechanisms: SE, RESIa, RESIb, and the contribution of all DI mechanisms combined.}
\label{pz1pz2}
\end{figure}

In \fig{pz1pz2} we show the correlated momenta of the two electrons for the SE, RESIa, RESIb pathways as well as the total pattern with all DI pathways combined. We find that the RESIa and RESIb pathways have distinct patterns
only for 10$^{14}$ and 1.5 $\times$ 10$^{14}$ Watts/cm$^2$ while for 3 $\times$ 10$^{14}$ Watts/cm$^{2}$ they are almost indistinguishable. For intensities in the tunneling regime the time electron 2 ionizes is around an extremum of the field for RESIa and a zero of the field for RESIb. However, in the over-the-barrier regime electron 2, in the Delayed pathway, ionizes mostly half way between an extremum and a zero of the field. Clearly, at the latter time the intense field lowers sufficiently the Coulomb potential to still allow electron 2 to escape.
Thus, in the following sections, we only present the combined contribution of RESIa and RESIb at 3 $\times$ 10$^{14}$ Watts/cm$^{2}$.

We further note that at 3 $\times$ 10$^{14}$ Watts/cm$^{2}$ the correlated momenta of RESI have a pronounced square structure, see \fig{pz1pz2}. This feature can be understood in terms of the
``soft" collisions that underly DI. For 70\% of the RESI trajectories the re-colliding electron ionizes at a very early time, before re-collision at T/2.
 For the latter trajectories the final momentum of the re-colliding electron is mostly determined by the vector potential at the time the electron tunnels out (first step in the three-step model). At 3 $\times$ 10$^{14}$ Watts/cm$^{2}$, electron 1 tunnels out at phases of the field centered around 50$^{\circ}$ and extending up to 60$^{\circ}$, see \fig{mean3013}. The momentum an electron gains by the field when the initial phase is around 50-60$^{\circ}$ is consistent
with the value of the momentum at the boundary of the square pattern of the RESI pathway. Since the RESI pathway prevails overall for strongly driven $N_{2}$, see following section, it is the square pattern that prevails in the combined 
correlated momenta, see \fig{pz1pz2}. Thus, {\it the correlated momenta probe the tunneling phase of the re-colliding electron for intermediate intensities in the over-the-barrier regime}. 
This is not the case in the tunneling regime; the two electrons undergo strong collisions and the final momenta for both electrons are determined by the vector potential at the re-collision time.


The SE correlated momenta have a pronounced anticorellation pattern as shown in \fig{pz1pz2}. For the majority of  SE trajectories the re-colliding electron tunnels at a phase around 50$^{\circ}$. For 30\% of the SE trajectories the re-colliding electron ionizes at an early time before re-collision at T/2. The second electron ionizes just after re-collision at T/2. Thus the two electrons ionize at times corresponding to opposite signs of the vector potential resulting in opposite final electron momenta. 
For the remaining 70\% of SE events both electrons ionize mostly after T/2.  Out of these 70\%, for some SE events the two electrons 
escape in the same direction, as is the case in the tunneling regime. However, for others, a small amount of energy transfer and possibly the Coulomb repulsion, result in the two electrons escaping in opposite directions. Overall electrons escaping in opposite directions is the biggest contribution to SE giving rise to an anticorrelation pattern.

\subsection {Double ionization probability distribution }
In this section we show that for  3 $\times$ 10$^{14}$ Watts/cm$^{2}$, in addition to the correlated momenta, the DI probability distribution as a function of total electron energy probes the tunneling phase of the re-colliding electron.

\begin{figure}[h] 
\centering 
\includegraphics [scale=0.32] {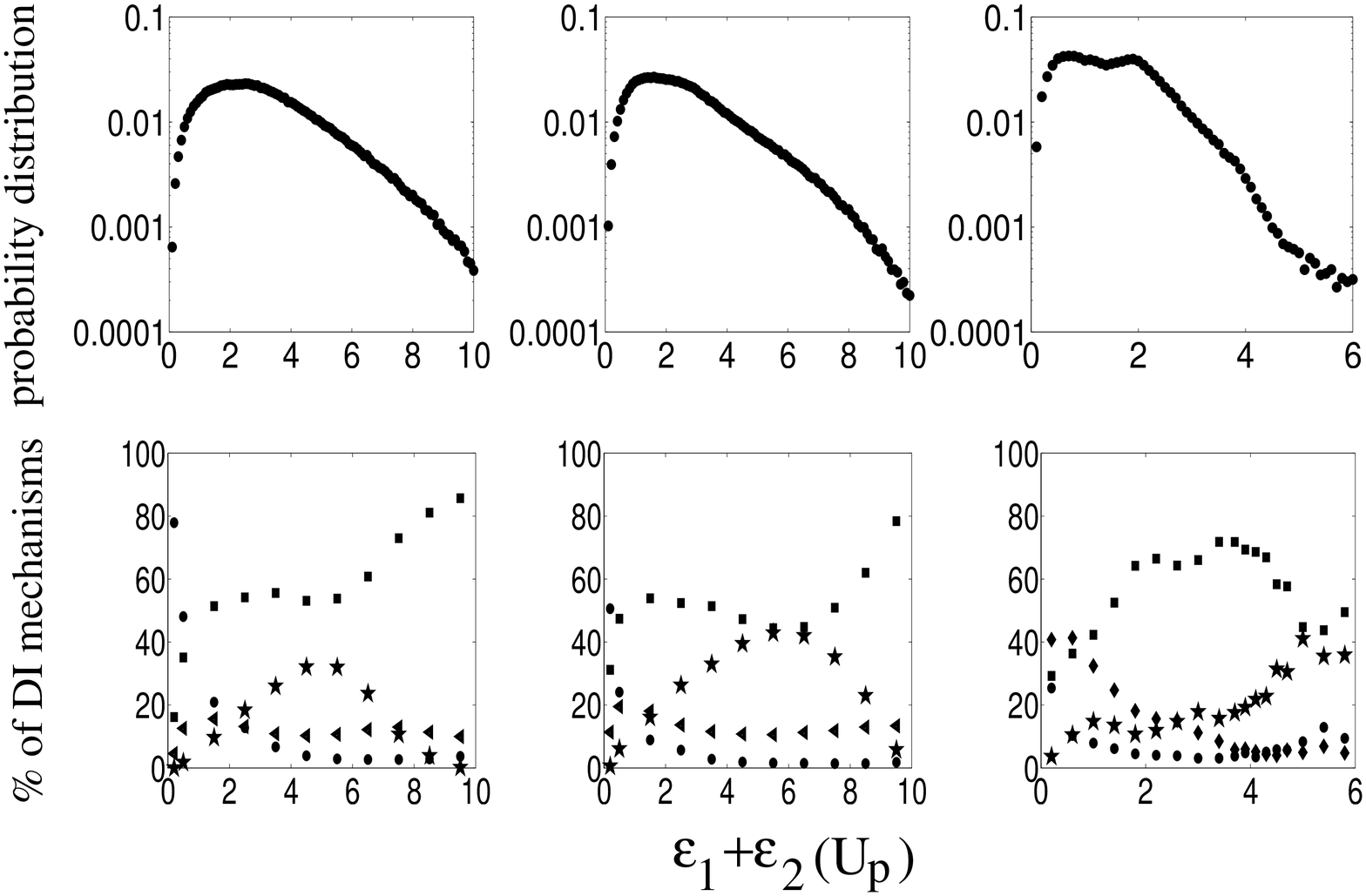} 
\caption{Top row: Double ionization probability distribution; Bottom row: \% contributions of the main DI mechanisms ($\bullet $ for DDE, $ \blacktriangleleft $ for RESIa, $ \blacksquare $ for RESIb, $ \bigstar$ for SE). First column: $10^{14}$ Watts/cm$^2$, second column: $1.5 \times 10^{14}$ Watts/cm$^2$, third column: $3 \times 10^{14}$ Watts/cm$^2$ (at this intensity an additional DI mechanism appears, the SI mechanism, given by $\blacklozenge$). Note that for $3 \times 10^{14}$ Watts/cm$^2$ we only show the combined contribution of RESIa and RESIb. } 
\label{mechanism}
\end{figure}

{\it a) \% contribution of DI pathways.}
In \fig{mechanism}, top row, we plot the probability distribution of DI for 10$^{14}$, 1.5 $\times$ 10$^{14}$ and 3$\times$ 10$^{14}$ Watts/cm$^2$. 
Moreover, for each intensity, we plot the \% contribution of each DI pathway as a function of the total electron energy, see bottom row in \fig{mechanism}. We use  energy steps as small as  the immense computational challenge of the endeavor allows---1-2 million DI events for the whole energy regime.

{\it Small total energies.}
 For small intensities the DDE pathway prevails for small total electron energy with a \%
contribution changing from 80\% for 10$^{14}$ Watts/cm$^2$ to 30\% for 3.0 $\times$ 10$^{14}$ Watts/cm$^{2}$. 
 We find that for $N_{2}$ in the DDE pathway the re-colliding electron does not get significantly trapped by the core after re-collision and thus the two electrons do not escape in opposite directions as is the case for strongly driven He, see \cite{EmmanouilidouPRA2010}. A possible explanation is that the presence of two nuclear centers instead of one with nuclear charge of one instead of two (He charge) makes it harder for the re-colliding electron to become bound after re-collision.  
 As the intensity increases to  3.0 $\times$ 10$^{14}$ Watts/cm$^{2}$ crossing over to the over-the-barrier regime,
 for small total energy, the sequential ionization (SI) pathway takes over. We define SI as the pathway where the two electrons are closest to each other at the beginning of the propagation. Thus the SI pathway is a signature of sequential ionization with both electrons ionizing mainly due to their interaction with the laser field. 

{\it Intermediate and high total energies.}
For all intensities considered, the Delayed pathway prevails for all energies, with the exception of very small ones. The \% contribution of SE is always smaller
than that of the RESI's. A trace of collisional physics on strong field ionization is the overall shape of the SE contribution to DI as a function of total energy. It is very small for small and high energies while it is large for intermediate energies, as we have already found for strongly driven He  \cite{EmmanouilidouPRA2010}. Thus, SE qualitatively resembles electron impact
ionization.  

 For strongly driven He, RESI prevails independent of the total electron energy \cite{EmmanouilidouNJP} for intensities where 3.2 U$_{p}$ is below the first excitation energy of He$^{+}$.  For higher intensities, for strongly driven He, the SE pathway prevails for intermediate total electron energy, see \cite{EmmanouilidouNJP, EmmanouilidouPRA2010}. In contrast, for strongly driven $N_{2}$, RESI prevails for all intensities. The larger contribution of RESI to strongly driven $N_{2}$, possibly, arises because the interaction of the field with the molecular ion makes available a much larger number of excited states to the bound electron compared to the atomic case. Upon return of the re-colliding electron to the molecular ion, electron 2 can be found in any of these excited states. It is thus less probable for electron 2 to be found close to either of the nuclei and to interact strongly with electron 1, escaping through SE; it is rather more probable for electron 2 to gain some energy from the re-colliding electron, get further excited, and subsequently ionize. 
 
 We also note, that the \% contribution of  
 different DI pathways depends on the duration of the pulse. Our finding in the current work  that RESI prevails for strongly driven $N_{2}$ stems from the fact that we are driving $N_{2}$ with a long laser pulse as is the case in our recent studies of strongly driven He \cite{EmmanouilidouNJP, EmmanouilidouPRA2010}.

{\it  b) DI probability distribution in the over-the-barrier regime.}   
For 3$\times$ 10$^{14}$ Watts/cm$^2$, we find that the DI probability distribution has two features that are not present for intensities in the tunneling regime. First the DI probability distribution has two peaks. Second the probability for two electrons to ionize with a large total energy is very small. \fig{mechanism} shows that for  3$\times$ 10$^{14}$ Watts/cm$^{2}$ the probability for two electrons to escape with a total energy of 6 U$_{p}$ is comparable to the probability for two electrons to escape with a total energy of 10U$_{p}$ when the intensity is 10$^{14}$ or 1.5$\times$ 10$^{14}$ Watts/cm$^2$. In what follows we explain the above two features.

\begin{figure} 
\centering 
\includegraphics [scale=0.3] {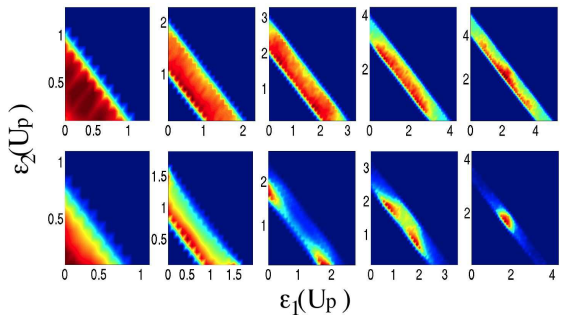} 
\caption{Double differential probability in energy: top row for 10$^{14}$ Watts/cm$^2$ for the energy intervals [0,1]U$_{p}$, [1,2]U$_{p}$, [2,3]U$_{p}$, [3,4]U$_{p}$ and [4,5]U$_{p}$; bottom row 
for $3 \times 10^{14}$ Watts/cm$^2$ for [0,0.8]U$_{p}$, [0.8,1.6]U$_{p}$, [1.6,2.4]U$_{p}$, [2.4,3.2]U$_{p}$ and [3.2,4]U$_{p}$.}
\label{shareenergy}
\end{figure}

As the intensity increases in the tunneling regime the peak of the DI probability distribution shifts to smaller total electron energy, from 2.5 U$_{p}$ at 10$^{14}$ Watts/cm$^2$
to 1.6 U$_{p}$ for 1.5$\times$10$^{14}$ Watts/cm$^2$, see \fig{mechanism}. We also find that around the maximum of the DI probability distribution the two electrons share the energy in all possible ways,
 see top row of \fig{shareenergy}. The electrons share the energy in a similar fashion for  3$\times$ 10$^{14}$ Watts/cm$^2$ when the total electron energy is around 0.7 U$_{p}$---the location of the first peak in the DI probability distribution. Thus, the peak that is present for all laser intensities considered shifts to smaller total energy with increasing intensity. This is consistent with a less efficient energy transfer from electron 1 to electron 2, case in point  the ``soft" collisions  that underly
 double ionization for the larger intensity considered in this study. The same reasoning holds for the overall shift of the DI probability distribution to smaller total energy.

\begin{figure} [h]
\centering 
\includegraphics [scale=0.45] {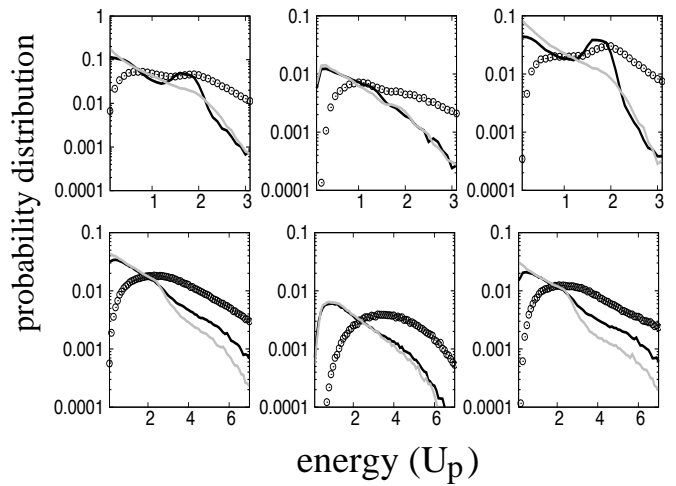} 
\caption{Top row $3 \times 10^{14}$ Watts/cm$^2$, first column from the left: total DI probability distribution as a function of total energy (circles) and probability distribution as a function of one electron's final energy, for the re-colliding electron (black line) and for the second electron (grey line). Second and third column same as the first but for the SE and RESI pathways. Bottom row, same as top row but for 10$^{14}$ Watts/cm$^2$.}
\label{mechanism1}
\end{figure}

We next explore the origin of the second peak around 1.9 U$_{p}$ for 3$\times$ 10$^{14}$ Watts/cm$^2$. From \fig{shareenergy} (bottom row) we find that, when the total energy is between 1.6-2.4 U$_{p}$, one electron---we find it to be the re-colliding one---escapes with most of the energy while electron 2 escapes with a very small part of it. This asymmetric energy sharing implies that the second peak in the total DI probability distribution is related to the re-colliding electron's probability distribution as a function of its final energy. We check whether this is the case by plotting, in \fig{mechanism1}, for each DI pathway the probability for each electron to escape with a certain amount of energy. We find that it is indeed the first electron's RESI probability distribution that accounts for the peak at higher total energy. Namely, in \fig{mechanism1}
we show that the RESI probability distribution of electron 1, as a function of the electron's final energy, varies smoothly for intensities in the tunneling regime (bottom row), in contrast, to its pronounced peak around 1.6 U$_{p}$ for $3 \times 10^{14}$ Watts/cm$^2$ (top row). To further understand the origin of this peak we 
also explore the dependence on total energy of the 70\% of RESI events where electron 1 ionizes very early on, before re-collision at T/2. We find that the \% contribution of the latter trajectories increases monotonically from 8\% at 0.4 U$_{p}$ total electron energy to its maximum value of 80\% for total energy at and above 1.6 $U_{p}$. We conclude that this second peak in the total DI probability distribution is directly related to these 70\% of RESI events. Indeed, the tunneling phase of 50-60$^{\circ}$ results in a final kinetic energy of 1.5 U$_{p}$ very close to the peak of the re-colliding electron's RESI probability distribution shown in \fig{mechanism1}. Thus this second peak in the DI probability distribution is yet another probe of the tunneling phase.

\section{Conclusions}
We have explored in detail the prevalence of the DI pathways for strongly driven $N_{2}$ for intensities ranging from the tunneling regime to intermediate ones in the over-the-barrier regime. We have found that the Delayed pathway prevails for all intensities and for all total energies, with the exception of very small energies. This is different from the atomic case where SE prevails for intermediate total energies for intensities where 3.2 U$_{p}$ is above the first excitation energy of He$^{+}$. 

For intermediate intensities in the over-the-barrier regime, we find that ``soft" collisions underly DI. In these ``soft'' collisions electron 1 tunnels at a large phase of the laser field, around 50$^{\circ}$, has a small maximum excursion in the laser field, and upon its return to the molecular ion transfers a small amount of energy to electron 2. One feature of ``soft" collisions is that electron 1 ionizes early on, before re-collision at T/2---this holds true for 70\% of RESI and for 30\% of SE trajectories.  For the  intermediate intensities in the over-the-barrier regime, we have shown that the boundaries of the square pattern of the correlated momenta as well as one of the two peaks of the DI probability distribution as a function of total energy probe the tunneling phase of the re-colliding electron.

\bibliographystyle{unsrt}

 \end{document}